\documentclass[12pt]{iopart}
\usepackage{epsfig,psfrag}
\usepackage{hyperref}

\def\nn{\nonumber}

\begin{document}

\title{Connecting the Chiral and Heavy Quark Limits : Full  Mass Dependence of Fermion Determinant in an Instanton Background}
\author{Gerald V. Dunne}
\address{Department of Physics, University of Connecticut, Storrs, CT 06269, USA}

\vskip 2.5 cm

\begin{abstract}
This talk reports work done in collaboration with Jin Hur, Choonkyu Lee and Hyunsoo Min concerning the computation of the precise mass dependence of the fermion determinant for quarks in the presence of an instanton background. The result interpolates smoothly between the previously known chiral and heavy quark limits of extreme small and large mass. The computational method makes use of the fact that the single instanton background has radial symmetry, so that the computation can be reduced to a sum over partial waves of logarithms of radial determinants, each of which can be computed numerically in an efficient manner using a theorem of Gelfand and Yaglom. The bare sum over partial waves is divergent and must be regulated and renormalized. We use the {\it angular momentum cutoff regularization and renormalization} scheme. Our results provide an extension of the Gelfand-Yaglom result to higher dimensional separable differential operators. I also comment on the application of this approach to a wide variety of fluctuation determinant computations in quantum field theory. 

\end{abstract}
\vskip 5cm
Talk presented at QFEXT05, Barcelona, September 2005.

\maketitle

\section{Introduction}
\label{introduction}

In this talk I present an overview of some recent progress \cite{idet} in computing one loop quantum vacuum polarization effects. 
Mathematically, this requires computing the determinant of a fluctuation operator, which describes the quadratic fluctuations about a semiclassical solution. This is a difficult problem, but it is worth studying as it has important physical applications to computations of the effective action, the partition function or the free energy. It is also an interesting mathematical problem to learn about the spectral properties of partial differential operators. In this talk I focus mainly on 
the computation of fermion determinants in nontrivial background
fields, which is an important challenge for both continuum and lattice
quantum field theory. Explicit analytic results are known only for
very simple backgrounds, and are essentially all variations on the
original work of Heisenberg and Euler
\cite{heisenberg,dunne}. For applications in
quantum chromodynamics (QCD), an important class of background gauge
fields are instanton fields, as these minimize the Euclidean gauge
action within a given topological sector of the gauge field.
Furthermore, instanton physics has many important phenomenological
consequences \cite{thooft,schaefer,shifman}. Thus,
we are led to consider the fermion determinant, and the associated
one-loop effective action, for quarks of mass $m$ in an instanton
background. Here, no exact results are known for the full mass
dependence, although several terms have been computed analytically
in the small mass \cite{thooft,carlitz,kwon} and large mass
\cite{nsvz,kwon} limits. Recently   \cite{idet}, with J. Hur, C. Lee and H.Min, we presented a new computation which is numerical, but
essentially exact, that evaluates the one-loop effective action in a
single instanton background, for any value of the quark mass
(and for arbitrary instanton size parameter). 
The result is fully consistent with the known small and large mass
limits, and interpolates smoothly between these limits. This could
be of interest for the extrapolation of lattice results
\cite{lattice}, obtained at unphysically large quark masses, to
lower physical masses, and for various instanton-based
phenomenology. 

Our computational method is simple and efficient, and
can be adapted to many other determinant computations in which the
background is sufficiently symmetric so that the problem can be
reduced to a product of one-dimensional radial determinants. While
this is still a very restricted set of background field
configurations, it contains many examples of interest, the single
instanton being one of the most obvious. 
The method is based on the Gelfand-Yaglom method for
computing determinants of \emph{ordinary} differential operators \cite{gy,levit,coleman,forman,kirsten,kleinertbook}. But in higher-dimensional problems with \emph{partial} differential
operators, it is known  \cite{forman} that the naive generalization is divergent, even for simple separable problems  where there is an infinite number of 1-D determinants to deal with. Physically, this divergence reflects the fact that in dimensions greater than one, 
one must confront  renormalization.
Our results can be viewed, in fact, as giving an extension of the Gelfand-Yaglom result to higher dimensional separable differential operators.

\section{Preliminaries:  Effective Action in an Instanton Background}
\label{renormalized}

The first step is to recall that since the instanton background field is self-dual, we can deduce the fermion determinant from a computation of the determinant of the associated Klein-Gordon operator. This is because self-dual gauge
fields have the remarkable property that the Dirac and Klein-Gordon
operators in such a background are isospectral
\cite{thooft,jackiw}. This implies that the
renormalized 
one-loop effective action of a Dirac spinor field of mass $m$ (and
isospin $\frac{1}{2}$), $\Gamma^{F}_{\rm ren}(A;m)$, is related to the corresponding scalar effective action, $\Gamma^{S}_{\rm ren}(A;m)$, 
for a complex scalar of mass $m$  (and isospin $\frac{1}{2}$) by
\cite{thooft, kwon}
\begin{equation}
\Gamma^{F}_{\rm ren}(A;m)= -2\, \Gamma^{S}_{\rm ren}(A;m)
-\frac{1}{2}\ln\left(\frac{m^2}{\mu^2}\right)\quad , \label{susy}
\end{equation}
where $\mu$ is the renormalization scale. The ln term in (\ref{susy}) corresponds to the existence of a zero
eigenvalue in the spectrum of the Dirac operator for a single
instanton background.  We consider an SU(2) single instanton background, in
regular gauge \cite{thooft,belavin} :
\begin{eqnarray}
\hskip -1cmA_{\mu}(x) \equiv A_{\mu}^{a}(x)\frac{\tau^{a}}{2}= \frac{\eta_{\mu\nu
a}\tau^{a}x_{\nu}}{r^2+\rho^2}\quad , \quad 
F_{\mu\nu}(x) \equiv F_{\mu\nu}^{a}(x)\frac{\tau^{a}}{2}
=-\frac{2\rho^2 \eta_{\mu\nu a}\tau^{a}}{(r^2+\rho^2)^2},
\label{instanton}
\end{eqnarray}
where $\eta_{\mu\nu a}$ are the standard 't Hooft symbols \cite{thooft,shifman}.

The one-loop effective action must be regularized. We use
Pauli-Villars regularization [with  heavy regulator mass $\Lambda$], adapted to the Schwinger proper-time
formalism, and later we relate this to dimensional regularization, as
in the work of 't Hooft \cite{thooft}. 
The regularized effective action has  the proper-time representation
\begin{eqnarray}
\hskip -2cm \Gamma_{\Lambda}^{S}(A;m) &=& - \int_{0}^{\infty}
\frac{ds}{s}(e^{-m^2 s}-e^{-\Lambda^2 s}) \int d^4
x\;\textrm{tr}\langle x|{e^{-s(-{\rm D}^2
)}-e^{-s(-\partial^2 )}}|x\rangle \quad ,
 \label{ptaction}
\end{eqnarray}
where $D^2 \equiv D_{\mu}D_{\mu}$, and 
$D_{\mu}=\partial_{\mu}-iA_{\mu}(x)$. The
renormalized effective action, in the minimal subtraction scheme,
is defined as \cite{thooft,kwon}
\begin{eqnarray}
\Gamma^{S}_{\rm ren}(A;m) &=& \lim_{\Lambda\rightarrow\infty}
\left[\Gamma_{\Lambda}^{S}(A;m)-\frac{1}{12} \frac{1}{(4\pi)^2}
\ln \left(\frac{\Lambda^2}{\mu^2}\right) \int d^4
x\;\textrm{tr}(F_{\mu\nu}F_{\mu\nu})\right] \nonumber \\
&\equiv& \lim_{\Lambda\rightarrow\infty}
\left[\Gamma_{\Lambda}^{S}(A;m)-\frac{1}{6}
\ln\left(\frac{\Lambda}{\mu}\right)\right]\quad ,
\label{renaction}
\end{eqnarray}
where we have subtracted the charge renormalization counter-term, and $\mu$ is the renormalization scale.
By dimensional considerations, we may
introduce the modified scalar effective action
$\tilde{\Gamma}^{S}_{\rm ren}(m\rho)$, which is a function of $m\rho$ only, defined by
\begin{equation}
\Gamma^{S}_{\rm ren}(A;m)=\tilde{\Gamma}^{S}_{\rm ren}(m\rho)+ \frac{1}{6}\ln(\mu\rho)\quad ,
\label{modaction}
\end{equation}
and concentrate on studying the $m\rho$ dependence of
$\tilde{\Gamma}^{S}_{\rm ren}(m\rho)$. Then there is
no loss of generality in our setting the instanton scale $\rho=1$ henceforth.

\begin{figure}[tp]
\psfrag{ver}{\Large $\tilde\Gamma_{\rm ren}(m)$}
\psfrag{hor}{\Large $m$}
\includegraphics[scale=1.7]{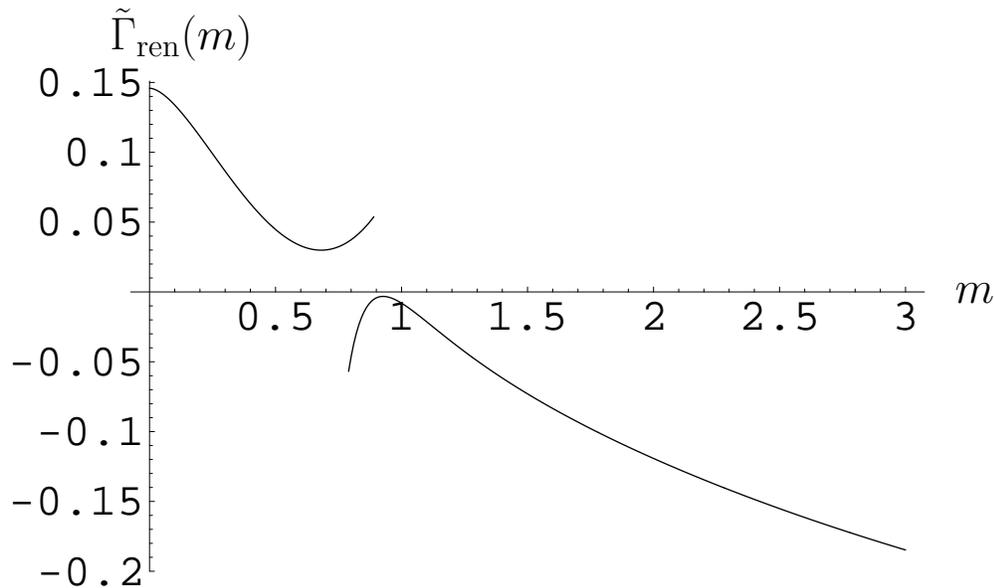}
\caption{Plot of the analytic small and large mass expansions for $\tilde{\Gamma}^{S}_{\rm ren}(m)$, from Equation (\protect{\ref{masslimit}}). Note the gap in the region $0.5\leq m \leq 1$, in which the two expansions do not match up.
\label{fig1}}
\end{figure}

It is known from previous work that in the small mass
\cite{thooft,carlitz,kwon} and large mass \cite{nsvz,kwon} limits,
$\tilde{\Gamma}^{S}_{\rm ren}(m)$ behaves as
\begin{eqnarray}
\hskip-2cm \tilde{\Gamma}^{S}_{\rm ren}(m)&=&
\begin{cases}
{\alpha(1/2)+\frac{1}{2}\left(\ln m+\gamma-\ln
2\right)m^2 +\dots \quad , \quad m\rightarrow 0 \cr 
\displaystyle {
 -\frac{\ln m}{6}-\frac{1}{75 m^2}-\frac{17}{735 m^4}+\frac{232}{2835 m^6}-\frac{7916}{148225 m^8}+\cdots \quad , \quad m\rightarrow \infty}}
\end{cases}
\label{masslimit}
\end{eqnarray}
where $\alpha(1/2)=-\frac{5}{72}-2 \zeta^\prime(-1)-\frac{1}{6}\ln 2  \simeq 0.145873$, and $\gamma\simeq 0.5772\dots$ is Euler's constant.  This small mass expansion is based on the fact that the
massless propagators in an instanton background are known in
closed-form. On the other hand, the large mass expansion
in (\ref{masslimit}) can be computed in several ways. 
A very direct approach is to use the small-$s$ behavior of 
the proper-time function appearing in (\ref{ptaction}), as given by
the Schwinger-DeWitt expansion.

Equation (\ref{masslimit}) summarizes what is known analytically about the mass dependence of the renormalized one-loop effective action in an instanton background. This situation is represented in Figure \ref{fig1}, which shows a distinct gap approximately in the region $0.5 \leq m\leq 1$, where the small and large mass expansions do not match up.

\section{The Gelfand-Yaglom Technique}
\label{gelfand}

A beautiful result of Gelfand and Yaglom \cite{gy} provides a spectacularly simple way to compute the determinant of a one-dimensional differential operator.
Their result has since been extended in various ways \cite{levit,coleman,forman,kirsten,kleinertbook}, but here we will only need their result for {\bf radial} differential operators.
Suppose ${\mathcal
M}_1$ and ${\mathcal M}_2$ are two second order ordinary
differential operators on the interval $r\in [0,\infty)$, with
Dirichlet boundary conditions assumed. In practice we will choose ${\mathcal
M}_1$ to be the operator of interest, and ${\mathcal
M}_2$ to be the corresponding {\it free} operator.  Then the ratio of the
determinants is given by
\begin{eqnarray}
\frac{\det {\mathcal M}_1}{\det{\mathcal M}_2}
=\lim_{R\to\infty}\left(\frac{\psi_1(R)}{\psi_2(R)}\right)
\label{theorem}
\end{eqnarray}
where $\psi_i(r)$ (for $i=1,2$ labelling the two different differential operators,  ${\mathcal
M}_1$ and ${\mathcal M}_2$) satisfies the {\bf initial value problem}
\begin{eqnarray}
{\mathcal M}_i\, \psi_i(r)=0 , \qquad {\rm with}\quad \psi_i(0)=0\quad {\rm and}\quad \psi_i^\prime(0)=1\quad .
\label{ode}
\end{eqnarray}
Note that no direct information about the spectrum
(either bound or continuum states, or phase shifts) is required in order to compute the determinant. All that is required is the integration of (\ref{ode}), which is straightforward to implement numerically.

\section{Radial Formulation}
\label{radial}

Returning now to the instanton determinant problem, we can use the fact that the single instanton background (\ref{instanton}) has radial symmetry \cite{thooft}. Thus, the regularized one-loop effective action (\ref{ptaction}) can be reduced to a sum over partial waves of logarithms of determinants of radial ordinary differential operators. Each such radial determinant can be computed using the Gelfand-Yaglom result (\ref{theorem}). Unfortunately, the sum over all partial waves is divergent. The physical challenge is to renormalize this divergent sum.

In the instanton background (\ref{instanton}), with scale $\rho=1$, the Klein-Gordon operator $-D^2$ for isospin $\frac{1}{2}$ particles can be cast in the radial form \cite{thooft}
\begin{equation}
\hskip -2cm -D^2\to  {\cal H}_{(l,j)} \equiv \left[ - \frac{\partial^2}{\partial
r^2}-\frac{3}{r}\frac{\partial}{\partial
r}+\frac{4l(l+1)}{r^2}+\frac{4(j-l)(j+l+1)}{r^2+1}-\frac{3}{(r^2+1)^2}
\right]\, ,
\label{insth}
\end{equation}
where $l=0,\; \frac{1}{2},\;1,\;\frac{3}{2},\;\cdots\;$, and $j=|\;l
\pm \frac{1}{2}\;|$, and there is a degeneracy factor of
$(2l+1)(2j+1)$ for each partial wave characterized by $(l,
j)$-values. In the absence of the instanton background, the
free operator is
\begin{equation}
-\partial^2\to {\cal H}^{\rm free}_{(l)} \equiv \left[ - \frac{\partial^2}{\partial
r^2}-\frac{3}{r}\frac{\partial}{\partial r}+\frac{4l(l+1)}{r^2}
\right] .
\label{freeh}
\end{equation}
This radial decomposition means that we can express the
Pauli-Villars regularized effective action also as
\begin{eqnarray}
\hskip -2cm \Gamma_\Lambda^S(A; m)  &=& \sum_{l=0,\frac{1}{2}, \dots} {\rm deg}(l) \left\{ \ln \left(\frac{\det [{\mathcal H}_{(l,l+\frac{1}{2})}+m^2]}{\det [{\mathcal H}^{{\rm free}}_{(l)}+m^2]}\right)+  \ln \left(\frac{\det[{\mathcal H}_{(l+\frac{1}{2},l)}+m^2]}{\det [{\mathcal H}^{{\rm free}}_{(l+\frac{1}{2})}+m^2]}\right)\right.
\nonumber\\
&&\hskip 2cm \left. - \ln \left(\frac{\det [{\mathcal H}_{(l,l+\frac{1}{2})}+\Lambda^2]}{\det [{\mathcal H}^{{\rm free}}_{(l)}+\Lambda^2]}\right)-  \ln \left(\frac{\det [{\mathcal H}_{(l+\frac{1}{2},l)}+\Lambda^2]}{\det [{\mathcal H}^{{\rm free}}_{(l+\frac{1}{2})}+\Lambda^2]}\right)\right\}
\label{pv}
\end{eqnarray}
Here we have combined the radial determinants for
$(l,j=l+\frac{1}{2})$ and
$(l+\frac{1}{2},j=(l+\frac{1}{2})-\frac{1}{2})$, which have the
common degeneracy factor ${\rm deg}(l)=(2l+1)(2l+2)$.

The Gelfand-Yaglom technique provides a simple and efficient numerical technique for computing each of the radial determinants appearing in (\ref{pv}). But to extract the renormalized effective action we need to be able to consider the $\Lambda\to\infty$ limit in conjunction with the infinite sum over $l$. This can be achieved as follows. Split the $l$ sum in (\ref{pv}) into two parts as :
\begin{eqnarray}
\Gamma_\Lambda^S(A; m)&=& \sum_{l=0,\frac{1}{2},\dots}^L \Gamma_{\Lambda, (l)}^S(A; m) + \sum_{l=L+\frac{1}{2}}^\infty \Gamma_{\Lambda, (l)}^S(A; m)
\label{actionsplit}
\end{eqnarray}
where $L$ is a large but finite integer. The first sum involves low partial wave modes, and the second sum involves the high partial wave modes. We consider each sum separately, before recombining them to obtain our final expression (\ref{answer2}).

\section{Low partial wave modes: determinants computed using Gelfand-Yaglom}

The first sum in (\ref{actionsplit}) is finite, so the cutoff  $\Lambda$ may be safely removed, and for each $l$ the determinant can be computed using the Gelfand-Yaglom result (\ref{theorem}).
We can simplify the numerical computation further by noting that for the
free massive Klein-Gordon partial-wave operator, ${\mathcal H}^{{\rm
free}}_{(l)}+m^2$ (with ${\mathcal H}^{{\rm free}}_{(l)}$ given in
(\ref{freeh})), the solution to (\ref{ode}) is 
\begin{eqnarray}
\psi^{\rm free}_{(l)}(r)=\frac{I_{2l+1}(m r)}{r}\quad .
\label{besseli}
\end{eqnarray}
This solution grows exponentially fast at large $r$, as do the
numerical solutions to (\ref{ode}) for the operators ${\mathcal
H}_{(l,j)}+m^2$, with ${\mathcal H}_{(l,j)}$ specified in
(\ref{insth}). Thus, it is numerically better to consider the ODE
satisfied by the {\it ratio} of the two functions
\begin{eqnarray}
{\mathcal R}_{(l,j)}(r)=\frac{\psi_{(l,j)}(r)}{\psi^{\rm free}_{(l)}(r)}\qquad ; \quad {\mathcal R}_{(l,j)}(0)=1 \quad ; \quad  {\mathcal R}^\prime_{(l,j)}(0)=0\quad .
\label{ratio}
\end{eqnarray}
 This quantity
has a finite value in the large $r$ limit, which is just the ratio of the determinants as in (\ref{theorem}).
In fact, since we are ultimately interested in the logarithm of the
determinant, it is more convenient (and more stable numerically) to
consider the logarithm of the ratio
 \begin{eqnarray}
 S_{(l,j)}(r)\equiv\ln {\mathcal R}_{(l,j)}(r)\quad ,
 \label{logfunction}
 \end{eqnarray}
which satisfies the differential equation
\begin{eqnarray}
\frac{d^2 S_{(l,j)}}{dr^2}+\left(\frac{d S_{(l,j)}}{dr}\right)^2+\left(\frac{1}{r}+2m\frac{I^\prime_{2l+1}(m r)}{I_{2l+1}(m r)}\right)\frac{d S_{(l,j)}}{dr}=U_{(l,j)}(r)\quad ,
\label{nonlinear}
\end{eqnarray}
with boundary conditions
\begin{eqnarray}
S_{(l,j)}(r=0)=0\qquad , \qquad S^\prime_{(l,j)}(r=0)=0\quad .
\label{logbc}
\end{eqnarray}
The `potential' term $U_{(l,j)}(r)$ in (\ref{nonlinear}) is given by
\begin{eqnarray}
U_{(l,j)}(r)=\frac{4(j-l)(j+l+1)}{r^2+1}-\frac{3}{(r^2+1)^2}.
\label{potential}
\end{eqnarray}
\begin{figure}[ht]
\psfrag{r}{\Large $r$}
\psfrag{l}{\Large $l$}
\psfrag{=}{\Large $\!=$}
\psfrag{0}{\Large $0$}
\psfrag{10}{\Large $10$}
\psfrag{20}{\Large $20$}
\psfrag{30}{\Large $30$}
\includegraphics[scale=1.7]{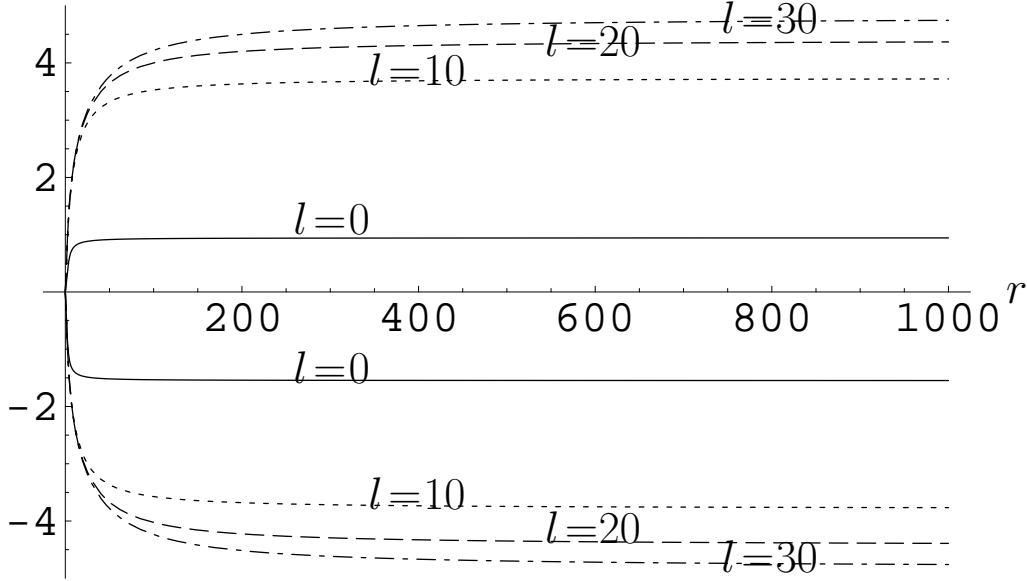}
\caption{Plots of the $r$ dependence of $S_{(l,l+\frac{1}{2})}(r)$
and $S_{(l+\frac{1}{2},l)}(r)$,
solutions of the nonlinear differential equation in (\protect{\ref{nonlinear}}),
for $m=1$, and for $l=0,10,20,30$.
The upper curves are for $S_{(l,l+\frac{1}{2})}(r)$, while the lower ones are for
$S_{(l+\frac{1}{2},l)}(r)$.
\label{fig2}}
\end{figure}

\begin{figure}[ht]
\psfrag{hor}{\Large $l$}
\psfrag{ver}{\Large $P(l)$}
\includegraphics[scale=1.5]{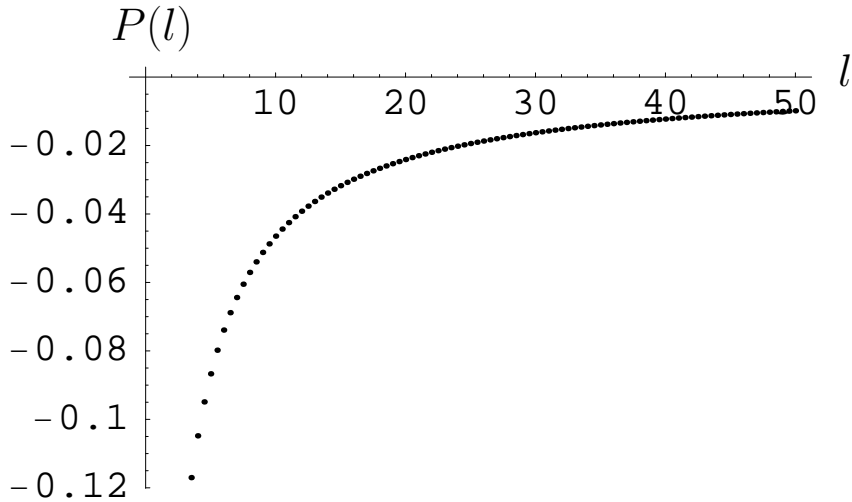}
\caption{Plot of the $l$ dependence of $P(l)=S_{(l,l+\frac{1}{2})}(r=\infty)+S_{(l+\frac{1}{2},l)}(r=\infty)$,  for $m=1$. $P(l)$ behaves like
$O(\frac{1}{l})$ for large $l$. 
\label{fig3}}
\end{figure}

To illustrate the computational method, in Figure \ref{fig2} we plot
$S_{(l,l+\frac{1}{2})}(r)$ and $S_{(l+\frac{1}{2},l)}(r)$ for
various values of $l$, with mass value $m=1$ (which is in the region
in which neither the large nor small mass expansions is accurate).
Note that the curves quickly reach an asymptotic large-$r$ 
constant value, and also notice that the contributions from
$S_{(l,l+\frac{1}{2})}(r=\infty)$ and
$S_{(l+\frac{1}{2},l)}(r=\infty)$ almost cancel one another when
summed. Indeed, 
\begin{eqnarray}
P(l)&\equiv&S_{(l,l+\frac{1}{2})}(r=\infty)+S_{(l+\frac{1}{2},l)}(r=\infty)\nonumber\\
& \sim& O \left( \frac{1}{l} \right)\qquad , \quad l\to\infty
\label{pl}
\end{eqnarray}
This behavior is illustrated in Figure \ref{fig3}. At first sight, this looks like bad news, because when including the degeneracy factor the sum over $l$
is
\begin{eqnarray} \sum_{l=0,\frac{1}{2},\dots}^L \Gamma_l^S(A;
m)&=&\sum_{l=0,\frac{1}{2}, \dots}^L (2l+1)(2l+2)  P(l) \label{lj1}
\end{eqnarray}
Therefore,  the sum (\ref{lj1}) has potentially divergent terms
going like $L^2$, $L$ and $\ln L$, as well as terms finite and
vanishing for large $L$. Remarkably, we find below that these divergent terms are
{\bf exactly canceled} by the divergent large $L$ terms found in the
previous section for the second sum in (\ref{actionsplit}).

\section{High partial wave modes: determinants computed using radial WKB}
In the second sum in (\ref{actionsplit}) we cannot take the large
$L$ and large $\Lambda$ limits blindly, as each leads to a
divergence. To treat these high $l$ modes, we use radial WKB \cite{wkbpaper}, which is a good approximation in precisely this limit. This means we can compute {\it analytically}  the large $\Lambda$ and
large $L$ divergences of the second sum in (\ref{actionsplit}), using the WKB approximation for the
corresponding determinants. This is a straightforward computation, the details of which can be found in \cite{idet,wkbpaper}. We find the following analytic expression for the  large $L$ behavior
\begin{eqnarray}
\hskip -2cm \sum_{l=L+\frac{1}{2}}^\infty \Gamma_{\Lambda, (l)}^S(A; m)&\sim& \frac{1}{6}\ln \Lambda+2 L^2 + 4 L-\left(\frac{1}{6}+\frac{m^2}{2}\right)\ln L \nonumber\\
&& +\left[\frac{127}{72}-\frac{1}{3}\ln 2+\frac{m^2}{2}-m^2 \ln 2+\frac{m^2}{2}\ln m \right]+O\left(\frac{1}{L}\right)
\label{div}
\end{eqnarray}
It is important to identify the physical role of the various terms in
(\ref{div}). The first term is the expected logarithmic counterterm
which is subtracted in (\ref{renaction}),
and explains the origin of the $\frac{1}{6}\ln \mu$ term in (\ref{modaction}).
The next three terms give
quadratic, linear and logarithmic divergences in $L$. We shall show
in the next section that these divergences cancel corresponding
divergences in the first sum in (\ref{actionsplit}), which were
found in our numerical data. It is a highly nontrivial check on this
WKB computation that these divergent terms have the correct
coefficients to cancel these divergences.
Note that the $\ln L$ coefficient, and the finite term, are mass dependent.

\section{Putting it all together}

\begin{figure}[tp]
\psfrag{hor}{\Large $m$}
\psfrag{ver}{\Large $\tilde\Gamma_{\rm ren}^S(m)$}
 \includegraphics[scale=1.75]{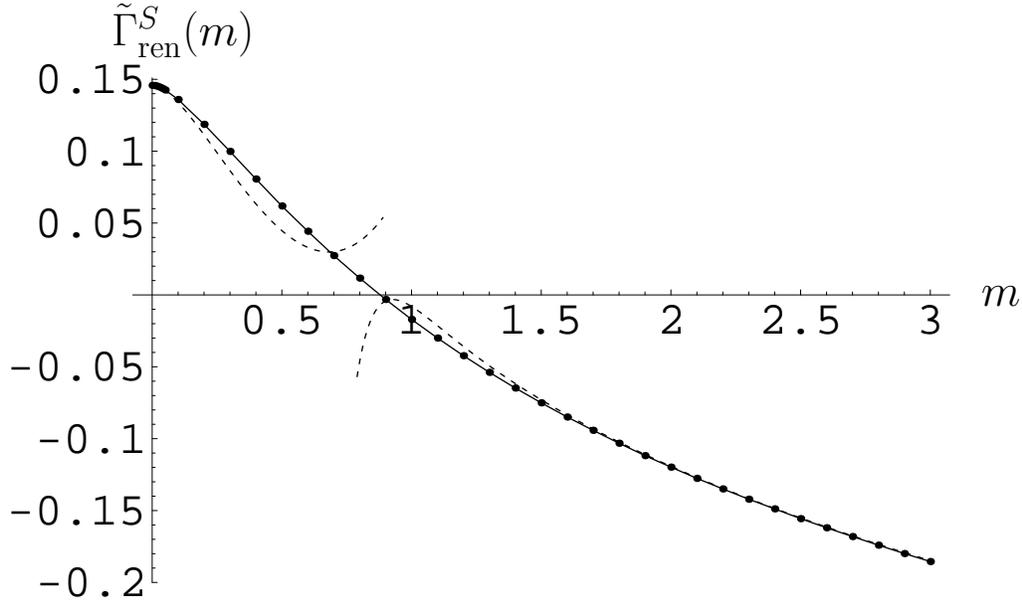}
\caption{Plot of our numerical results for $\tilde{\Gamma}^{S}_{\rm ren}(m)$
from (\protect{\ref{answer2}}), compared  with the analytic extreme small
and large mass limits
[dashed curves] from (\protect{\ref{masslimit}}).
The dots denote numerical data points from (\protect{\ref{answer2}}),
and the solid line is a fit through these points. The agreement with the analytic small and large mass limits is very precise.
\label{fig4}}
\end{figure}

We now combine the numerical results for the low partial wave modes with the radial WKB results for the high partial wave modes to obtain the minimally subtracted renormalized effective
action $\tilde{\Gamma}^S_{\rm ren}(m)$ as
\begin{eqnarray}
\hskip-1cm \tilde{\Gamma}^S_{\rm ren}(m)&=&\lim_{L\to \infty}\left\{\sum_{l=0,\frac{1}{2},\dots}^L (2l+1)(2l+2) P(l)
+2 L^2 + 4 L-\left(\frac{1}{6}+\frac{m^2}{2}\right)\ln L  \right.\nonumber \\
&&\left. +\left[\frac{127}{72}-\frac{1}{3}\ln 2+\frac{m^2}{2}-m^2
\ln 2+\frac{m^2}{2}\ln m \right]\right\}.
\label{answer2}
\end{eqnarray}
The remarkable observation is that the large L divergences found in the numerical results are precisely cancelled by the analytic large L divergence found using WKB, leaving a finite answer for the renormalized effective action. This holds for any mass $m$. In Figure
\ref{fig4} we plot these results for $\tilde{\Gamma}^{S}_{\rm ren}(m)$,
and compare them with the analytic
small and large mass expansions in (\ref{masslimit}). The agreement
is spectacular. Thus, our expression (\ref{answer2}) provides a
simple and numerically precise interpolation between the large mass
and small mass regimes.

As an interesting analytic check, our formula (\ref{answer2}) provides a very
simple computation of 't Hooft's leading small mass result. When $m=0$, the $P(l)$ can be computed analytically:
$P(l)=\ln\left[\frac{2l+1}{2l+2}\right]$. Then
\begin{eqnarray}
\hskip -2cm \tilde{\Gamma}^S_{\rm ren}  &=&\lim_{L\to\infty}\left\{\sum_{l=0,\frac{1}{2},\dots}^L {\rm deg}(l)
\ln\left(\frac{2l+1}{2l+2}\right) +2 L^2 + 4 L-\frac{1}{6}\ln L
+\frac{127}{72}-\frac{1}{3}\ln 2\right\} \nonumber\\
&=& -\frac{17}{72}-\frac{1}{6}\ln 2 +\frac{1}{6}-2\zeta^\prime(-1)\nonumber\\
&=& \alpha\left(\frac{1}{2}\right)= 0.145873...
\label{masslessanswer}
\end{eqnarray}
which agrees precisely with the leading term in the small mass limit in (\ref{masslimit}).

\section{Comparison With The Derivative Expansion}
\label{comparison}

\begin{figure}[tp]
\psfrag{hor}{\Large $m$}
\psfrag{ver}{\Large $\tilde\Gamma_{\rm ren}^S(m)$}
 \includegraphics[scale=1.75]{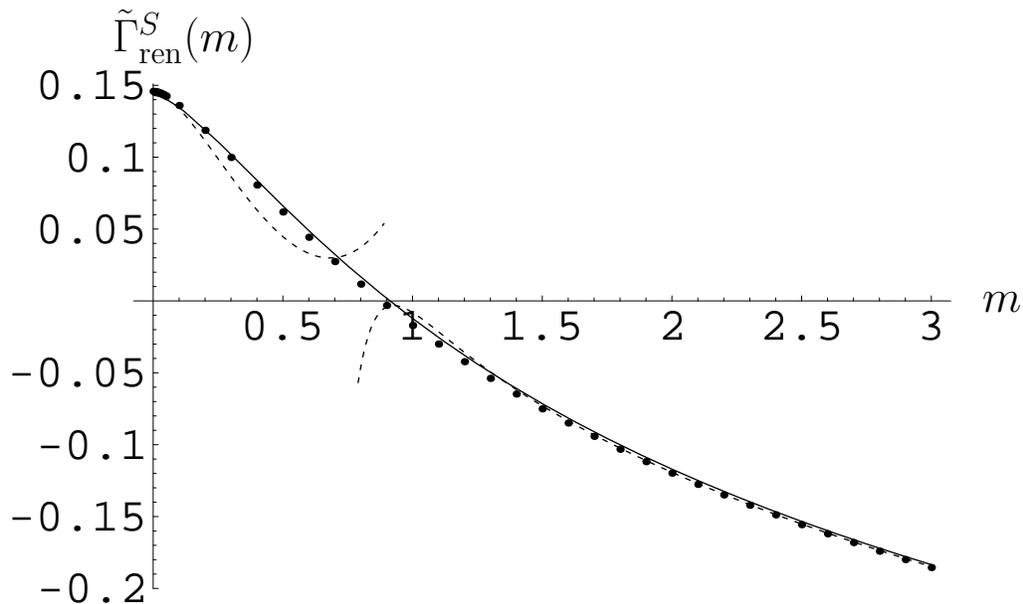}
\caption{Plot of $\tilde{\Gamma}^{S}(m)$, comparing the leading derivative expansion
approximation (solid line) from (\ref{dm}) with the precise numerical answers (dots). The dashed lines show the small and large mass limits from (\protect{\ref{masslimit}}).
\label{fig5}}
\end{figure}

In \cite{wkbpaper}, the renormalized effective action $\tilde{\Gamma}^{S}_{\rm ren}(m)$ was computed using the derivative expansion. Recall that the philosophy of the
derivative expansion is to compute the one-loop effective action for
a covariantly constant background field, which can be done exactly,
and then perturb around this constant background solution. The
{\bf leading order} derivative expansion for the effective action is
obtained by first taking the (exact) expression for the effective
Lagrangian in a covariantly constant background, substituting the
space-time dependent background, and then integrating over
space-time. For an instanton background, which is self-dual, we base
our derivative expansion approximation on a covariantly constant and
self-dual background \cite{dunne, ds1}. 
This leads to the following simple integral representation for the
leading derivative expansion approximation to the effective action
\cite{wkbpaper}
\begin{eqnarray}
\hskip -2cm \left.\tilde{\Gamma}^S_{\rm ren}(A; m)\right]_{\rm DE}
&=&-\frac{1}{14}\int_0^\infty \frac{dx\, x}{e^{2\pi x}-1}\left\{ 14 \ln\left(1+\frac{48 x^2}{m^4}\right)
+7\sqrt{3}\,\frac{m^2}{x} {\rm arctan}\left(\frac{4\sqrt{3}\, x}{m^2}\right) \right.\nn\\
&&\left. \hskip 1cm -84+768\frac{x^2}{m^4}~_2 F_1\left(1,\frac{7}{4},\frac{11}{4};-\frac{48x^2}{m^4}\right)\right\} -\frac{1}{6}\ln m
\label{dm}
\end{eqnarray}
Figure \ref{fig5} shows a comparison of this leading derivative
expansion expression with the exact numerical data. In the range
covered, the agreement is surprisingly good for such a crude
approximation. One would expect good agreement in the large $m$ limit, but not in the small $m$ limit. I conjecture that this remarkable agreement for both small and large mass is due to the special conformal symmetry of the single instanton background \cite{thooft,jackiw}. This is physically sensible, but has not yet been proved.

\section{Concluding Comments}
\label{conclusion}

Our {\it angular momentum cutoff regularization and renormalization} procedure of using the Gelfand-Yaglom result numerically for the low partial wave modes, and radial WKB for the exact large $L$ behavior for the high partial wave modes, produces a finite renormalized effective action which interpolates precisely between the previously known small and large mass limits. This is a highly nontrivial check, as all three computations (small mass \cite{thooft}, large mass \cite{nsvz,kwon}, and all mass \cite{idet}) are independent. Our result also provides a simple interpolation formula \cite{idet} which can be translated into the instanton scale dependence at fixed quark mass, and this may be of phenomenological use.

Our computational method is versatile and can be
adapted to a large class of previously insoluble computations of
one-loop functional determinants in nontrivial backgrounds in
various dimensions of space-time, as long as the spectral problem of
the given system can be reduced to that of partial waves. Hyunsoo Min's talk at this conference reports a new computation \cite{fv} of the false vacuum decay rate in a four dimensional self-interacting scalar field theory. Here the classical "bounce" solution is radial in 4d Euclidean space, so our technique is ideally suited to the computation of the fluctuation determinant about this bounce. An interesting feature here is the presence of a negative mode and zero modes. A different regularization method, also based on Gelfand-Yaglom, has  been applied to the false vacuum decay problem by Baacke and Lavrelashvili \cite{baacke}. See also the very interesting earlier work of Isidori et al \cite{isidori}.

Finally, we note that our method provides an extension of the Gelfand-Yaglom result for ODE's to the case of separable PDE's. As noted by Forman \cite{forman} for the 2d radial disc problem, the naive extension does not work because the product over angular momenta diverges, even though the result for each angular momentum is finite. Physically, this is because in higher dimensions renormalization is required. Our method incorporates renormalization and yields a finite physical answer for the determinant of the separable partial differential operator.

\vskip .25cm {\bf Acknowledgments:} This talk is based on work done with J. Hur, C. Lee and H. Min, whom I thank for a fruitful and enjoyable collaboration. I also thank H. Gies,  V. Khemani
and K. Kirsten for helpful comments, and the US DOE for support through
the grant DE-FG02-92ER40716. Finally, I heartily thank the QFEXT05 organizers, especially Emilio Elizalde, for an inspiring conference in a truly inspiring city!


\vskip .25cm 


\begin{thebibliography}{12345}

\bibitem{idet}
  G.~V.~Dunne, J.~Hur, C.~Lee and H.~Min,
  ``Precise quark mass dependence of instanton determinant,''
  Phys.\ Rev.\ Lett.\  {\bf 94}, 072001 (2005)
  [arXiv:hep-th/0410190];
``Calculation of QCD instanton determinant with arbitrary mass,''
  Phys.\ Rev.\ D {\bf 71}, 085019 (2005)
  [arXiv:hep-th/0502087].

\bibitem{heisenberg}
W.~Heisenberg and H.~Euler,
``Consequences Of Dirac's Theory Of Positrons,''
Z.\ Phys.\  {\bf 98}, 714 (1936);
V. Weisskopf, 
``The electrodynamics of the vacuum based on the quantum theory of the
electron,''
Kong. Dans. Vid. Selsk. Math-fys. Medd. XIV No. 6 (1936);
J.~S.~Schwinger,
``On Gauge Invariance And Vacuum Polarization,''
Phys.\ Rev.\  {\bf 82}, 664 (1951).

\bibitem{dunne} For a recent review, see: G.~V.~Dunne,
``Heisenberg-Euler effective Lagrangians: Basics and extensions,''
arXiv:hep-th/0406216,
In Ian Kogan Memorial Collection, {\it From Fields to Strings: Circumnavigating Theoretical Physics}, M. Shifman et al (Eds), (World Scientific, Singapore, 2004), Volume I, pages 445-522.

\bibitem{thooft}
G.~'t Hooft,
``Computation Of The Quantum Effects Due To A Four-Dimensional
Pseudoparticle,''
Phys.\ Rev.\ D {\bf 14}, 3432 (1976)
[Erratum-ibid.\ D {\bf 18}, 2199 (1978)].

\bibitem{schaefer}
T.~Schafer and E.~V.~Shuryak,
``Instantons in QCD,''
Rev.\ Mod.\ Phys.\  {\bf 70}, 323 (1998)
[arXiv:hep-ph/9610451].

\bibitem{shifman}
M.~A.~Shifman,
{\it ITEP lectures on particle physics and field theory.  Vol. 1 and 2},
World Sci.\ Lect.\ Notes Phys.\  {\bf 62}, 1 (1999).


\bibitem{carlitz} 
R.~D.~Carlitz and D.~B.~Creamer,
``Light Quarks And Instantons,''
Ann. Phys.\  {\bf 118}, 429 (1979).

\bibitem{kwon}
O.~K.~Kwon, C.~Lee and H.~Min, 
``Massive field contributions to the QCD vacuum tunneling amplitude,''
Phys.\ Rev.\ D {\bf 62}, 114022 (2000)
[arXiv:hep-ph/0008028].

\bibitem{nsvz}
V.~A.~Novikov, M.~A.~Shifman, A.~I.~Vainshtein and V.~I.~Zakharov,
``Calculations In External Fields In Quantum Chromodynamics'', 
Fortsch.\ Phys.\  {\bf 32}, 585 (1985).


\bibitem{lattice}
A.~W.~Thomas, ``Chiral Extrapolation of Hadronic Observables'', 
 Nucl.\ Phys.\ Proc.\ Suppl.\  {\bf 119}, 50 (2003)
[arXiv:hep-lat/0208023]; 
C.~Bernard et al, ``Panel discussion on chiral extrapolation of physical observables'', 
Nucl.\ Phys.\ Proc. \ Suppl.\  {\bf 119}, 170 (2003) [arXiv:hep-lat/0209086].

\bibitem{gy}
  I.~M.~Gelfand and A.~M.~Yaglom,
  ``Integration In Functional Spaces And It Applications In Quantum Physics,''
  J.\ Math.\ Phys.\  {\bf 1}, 48 (1960).

\bibitem{levit}
S. Levit and U. Smilansky, ``A Theorem on Infinite Products of Eigenvalues of Sturm-Liouville Operators'', Proc. Am. Math. Soc. {\bf 65}, 299 (1977).

\bibitem{coleman}
S. Coleman, {\it The Uses of Instantons}, Erice Lectures 1977, reprinted in S. Coleman, {\it Aspects of Symmetry}, (Cambridge University Press, 1988).


\bibitem{forman}
R. Forman, ``Functional Determinants and Geometry'', Invent. Math. {\bf 88}, 447 (1987).

\bibitem{kirsten}
K.~Kirsten and A.~J.~McKane,
``Functional determinants by contour integration methods,''
Ann. Phys. (NY) {\bf 308}, 502 (2003),
[arXiv:math-ph/0305010];
``Functional determinants for general Sturm-Liouville problems,''
J. Phys. A. {\bf 37}, 4649  (2004),
[arXiv:math-ph/0403050];
K. Kirsten, {\it Spectral Functions in Mathematics and Physics}, (Chapman-Hall, Boca Raton, 2002).

\bibitem{kleinertbook}
  H.~Kleinert,
  ``Path Integrals in Quantum Mechanics,  Statistics, Polymer Physics,
  and Financial Markets,'' (World Scientific, Singapore, 2004).



\bibitem{jackiw}
A.~S.~Schwarz,
``On Regular Solutions Of Euclidean Yang-Mills Equations,''
Phys.\ Lett.\ B {\bf 67}, 172 (1977);
L.~S.~Brown, R.~D.~Carlitz and C.~Lee,
``Massless Excitations In Instanton Fields,''
Phys.\ Rev.\ D {\bf 16}, 417 (1977);
R.~Jackiw and C.~Rebbi,
``Spinor Analysis Of Yang-Mills Theory,''
Phys.\ Rev.\ D {\bf 16}, 1052 (1977);

\bibitem{belavin}
A.~A.~Belavin, A.~M.~Polyakov, A.~S.~Shvarts and Y.~S.~Tyupkin,
``Pseudoparticle Solutions Of The Yang-Mills Equations,''
Phys.\ Lett.\ B {\bf 59}, 85 (1975).


\bibitem{wkbpaper}
G.~V.~Dunne, J.~Hur, C.~Lee and H.~Min, 
``Instanton determinant with arbitrary quark mass: WKB phase-shift method and
derivative expansion,''
Phys.\ Lett.\ B {\bf 600}, 302 (2004)
[arXiv:hep-th/0407222].


\bibitem{ds1}
G.~V.~Dunne and C.~Schubert,
``Two-loop self-dual Euler-Heisenberg Lagrangians. I: Real part and  helicity
amplitudes,''
JHEP {\bf 0208}, 053 (2002)
[arXiv:hep-th/0205004].


\bibitem{fv} G.~V.~Dunne and H.~Min, ``Beyond the thin-wall approximation : precise numerical computation of prefactors in false vacuum decay'', [arXiv:hep-th/0511156]. Also see H. Min, talk at QFEXT05, Barcelona, September 2005.

\bibitem{baacke}
  J.~Baacke and G.~Lavrelashvili,
  ``One-loop corrections to the metastable vacuum decay,''
  Phys.\ Rev.\ D {\bf 69}, 025009 (2004)
  [arXiv:hep-th/0307202].
  
  \bibitem{isidori}
  G.~Isidori, G.~Ridolfi and A.~Strumia,
  ``On the metastability of the standard model vacuum,''
  Nucl.\ Phys.\ B {\bf 609}, 387 (2001)
  [arXiv:hep-ph/0104016].


\end{thebibliography}
\end{document}